\documentclass[thmsa,12pt]{article}
\usepackage{amssymb}
\usepackage{graphicx}

\usepackage{amsmath}

\begin{document}

\title{Generation of polarization entangled photon pairs by a single crystal
interferometric source pumped by femtosecond laser pulses}
\author{M. Barbieri, C. Cinelli, F. De Martini, P. Mataloni \\
Dipartimento di Fisica, Universit\`{a} di Roma ''La Sapienza'' \\
and Consorzio Interuniversitario \\
per le Scienze Fisiche della Materia, \\
Roma, 00185 ITALY\\
e-mail: paolo.mataloni@uniroma1.it}
\date{}
\maketitle

\begin{abstract}
Photon pairs, highly entangled in polarization have been generated under
femtosecond laser pulse excitation by a type I crystal source, operating in
a single arm interferometric scheme. The relevant effects of temporal
walk-off existing in these conditions between the ordinary and extraordinary
photons were experimentally investigated. By introducing a suitable temporal
compensation between the two orthogonal polarization components highly
entangled pulsed states were obtained.
\end{abstract}

\section{Introduction}

Quantum states of polarization entangled photons, generated by the
Spontaneous Down Conversion (SPDC) process, are widely used for the
realization of important tasks of quantum information and are proposed for
several schemes of quantum cryptography \cite{1,2,3}. Pairs of correlated
SPDC photons at wavelenght (wl) $\lambda _{A}$ and $\lambda _{B}$ and
wavevector $\mathbf{k}_{A}$ and $\mathbf{k}_{B}$ are generated by a
nonlinear (NL) optical crystal shined by a pump leaser beam at wl $\lambda
_{p}$ and wv $\mathbf{k}_{p}$. In presence of polarization, $\pi -$%
entanglement and degenerate parametric emission ($\lambda _{A}=\lambda
_{B}=\lambda =\lambda _{p}/2$ ) the bipartite $2\times 2$ Hilbert space is
spanned by the Bell state basis:
\begin{eqnarray}
\left| \Psi ^{\pm }\right\rangle &=&\frac{1}{\sqrt{2}}\left( \left|
H_{A}V_{B}\right\rangle \pm \left| V_{A}H_{B}\right\rangle \right) , \\
\left| \Phi ^{\pm }\right\rangle &=&\frac{1}{\sqrt{2}}\left( \left|
H_{A}H_{B}\right\rangle \pm \left| V_{A}V_{B}\right\rangle \right) ,
\end{eqnarray}
expressed in the horizontal ($H$) and vertical ($V$) polarization basis.

Generally, polarization, $\pi -$entangled photon states are generated by a
type II noncollinear ''phase matched'' nonlinear (NL) crystal in which a
pair of mutually orthogonally polarized photons is generated over two
particular correlated $\mathbf{k}$ modes \cite{4}. The overall quantum state
emitted over these two modes is generally expressed by either Bell state $%
|\Psi ^{\pm }\rangle $, depending on a preset NL\ crystal orientation. Other
sources of $\pi -$entangled pairs are based on SPDC radiation generated by
coherently pumping two thin type I crystals, orthogonally oriented and
placed in mutual contact \cite{5} and by the interferometric parametric
source, in which a single type I crystal is excited in two opposite
directions ($\mathbf{k}_{p}$, $-\mathbf{k}_{p}$) by a back-reflected pump
beam \cite{6}. In both cases the spatial indistinguishability of the two
parametric emission cones of the crystals, which are associated to the
orthogonal polarizations, $H$ and $V$, are responsible of the polarization
entanglement corresponding to the Bell states $|\Phi ^{\pm }\rangle $.
Several examples of two photon state manipulation, allowing the generation
of either pure or mixed states, with variable degree of entanglement and
mixedness, have been realized \cite{7,8} by these sources. More recently,
their flexibility has been demonstrated in the generation of hyper-entangled
photon pairs \cite{9,10}.

A great deal of effort has been devoted towards development of sources of
entangled photons operating under \textit{fsec} laser pulse pumping. While this
is generally done with type II phase matching \cite{11}, only few examples
of pulsed entangled state generation have been given for a double crystal
system operating in the \textit{fsec} regime \cite{12}. It is well known that in
this case temporal walk-off may determine a strong degradation of $\pi -$%
entanglement because of the distinguishability introduced on the photon
path-ways by the effects of dispersion and birefringence in the NL crystal
\cite{12, 13}. In the present paper we report on the efficient generation of
polarization entangled pairs, obtained when the interferometric single
crystal scheme \cite{6} operates in this temporal regime. The effects of
relevant walk-off affecting the entangled state generation in this
particular case have been considered. Moreover, we have performed a detailed
investigation of the properties of entanglement and nonlocality of the
parametric source and compared the experimental results with the theoretical
predictions.

\section{The single crystal interferometric source}

\subsection{Description}

Let's briefly describe the parametric source shown in Figure 1. It is based
on a \textit{single arm} interferometer whose active element is given by a
single type-I, $l=0.5mm$ thick, $\beta $-barium-borate (BBO) crystal, cut at
the phase matching angle $\vartheta =29.3^{\circ }$. It is excited in two
opposite directions, $\mathbf{k}_{p}$ and $-\mathbf{k}_{p}$, by a single
transverse mode, vertically ($V$) polarized, $2^{nd}$ harmonic beam of a
Ti-Sapphire (Coherent, MIRA $900$) $f\sec $ laser ($\lambda _{p}=397.5nm$).
The pump beam is slightly focused by a $f=150$ $cm$ lens on the BBO. $H$
polarized degenerate photon pairs are generated at $\lambda =795$ $nm$ over
an emission cone with full aperture $\simeq $ $6%
{{}^\circ}%
$. The interferometric scheme performs the simultaneous reflection by a
common spherical mirror $M$ of either the pump beam or the parametric
radiation. It includes also a $\lambda $ quarter wave plate ($\lambda -$qwp)
which plays a fundamental role in the generation of the entangled state by
performing the $H\rightarrow V$ transformation for the SPDC radiation. This
scheme is the same of that previously described in the case of cw pumping
operation \cite{6}. After transmission through a $f=15cm$ lens ($L$) the
photons belonging to the emission cone are transmitted through a circular
mask which identifies the so-called \textit{entanglement-ring} ($e-ring$).
More than $15\cdot 10^{3}$ photon pairs corresponding to the $\pi -$%
entangled state $\left| \Phi \right\rangle =\frac{1}{\sqrt{2}}(\left|
H_{A}H_{B}\right\rangle +e^{i\phi }\left| V_{A}V_{B}\right\rangle )$ are
detected for a pump power of $150mW$, within the spectral bandwidth $\Delta
\lambda =3nm$ of the interference filters ($IF$), which corresponds to a
coherence-time of the emitted photons:\ $\tau _{coh}$\textit{\ }$\approx
447f\sec $. It is worth noting that a very high phase stability of the state
is guaranteed by the particular configuration of single arm interferometer
adopted for the source. Indeed, the triplet ($\phi =0$) to singlet ($\phi
=\pi $) state transition is performed by a macroscopic displacement ($\simeq
70$ $\mu m$) of mirror $M$ with respect the BBO crystal.
\begin{figure}[t]
\label{source}
\includegraphics[scale=0.5]{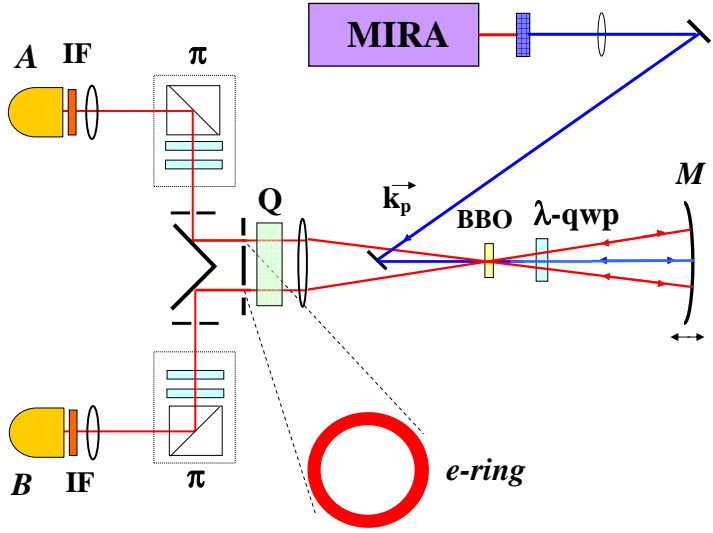}
\caption{Layout of the source of polarization entangled photon
states operating in pulsed regime.}
\end{figure}

A significant test of the spatial and temporal superposition of the two
cones, in absence of polarization entanglement (i.e. with the $\lambda -$qwp
oriented at $0%
{{}^\circ}%
$ with respect its optical axis), is given by the measurement of the
interference visibility performed by changing the phase $\phi $\ and
detecting an increasing fraction of the $e-ring$. This condition corresponds
to the absence of any temporal or spatial walk-off between the photon pairs
which can be generated with equal probability either in the first or in the
second passage through the crystal. The single and the coincidence rates
vary in this case proportionaly to $\cos ^{2}(\frac{\theta }{2})$. We could
measure $V\geq 0.9$ over the entire $e-ring$, while a maximum of $V=0.93$
was obtained for an arc length of $\simeq 2mm$.

\subsection{Temporal and spatial walk-off}

The two birefringent (b.r.) elements, represented by the BBO crystal and the
$\lambda -$qwp, may determine a relevant walk-off between the $H$ and $V$
polarization components when polarization entanglement is performed (i.e.
with $\lambda -$qwp oriented at $45%
{{}^\circ}%
$). In particular, when passing through a b.r. medium of length $x$, the
temporal walk-off occurring between the ordinary and extraordinary
wavepackets, with group velocities $v_{o}$ and $v_{e}$, is
\begin{equation*}
\tau =x\left( \frac{1}{v_{o}}-\frac{1}{v_{e}}\right)
\end{equation*}
We need to consider this effect for each b.r. element present inside the
parametric source (in the present case the BBO and the $\lambda -$qwp) and
compare the resultant value of $\tau $ with the coherence-time $\tau _{coh}$.

Spatial walk-off arises from the different output angles at which the two
waves go out of each b.r. element. This corresponds to a mutual spatial
displacement $\delta $ of the two beams, given by
\begin{equation*}
\delta =x\tan \alpha \thicksim x\alpha \text{,}
\end{equation*}
where $\alpha $ represents the mutual angle between the two waves. Spatial
walk-off becomes particularly severe when $\delta $ is of the order of the
beam diameter.
\begin{figure}[t]
\centering
\label{fig2}
\includegraphics[scale=0.6]{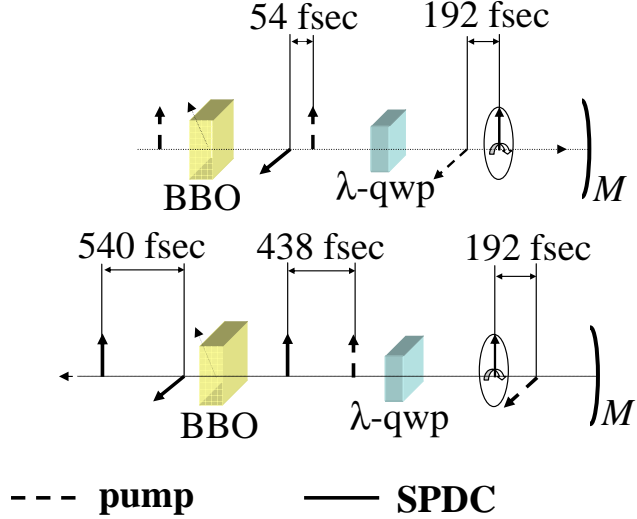}
\caption{Schematic picture of the temporal delays occurring between
the pump and the $H$ and $V$ polarized photons in each section of the
parametric source.}
\end{figure}

Let's evaluate first the temporal walk-off effect. Here we assume that any $%
H $ polarized photon pair generation event occurs in the center $l/2$ of
BBO. Moreover, either the pump pulse or the photon pair generated in the
first passage passes two times with their own polarizations through the NL
crystal and the zero-order $\lambda -$qwp. It is made by putting together
two hortogonally oriented quartz crystals, with thickness $l^{\prime }$ and $%
l^{\prime }+\lambda /4$, where $l^{\prime }=1.8mm$ \cite{14}. The total
temporal walk-off effect can be evaluated by looking at the scheme of Figure
2 which shows how the temporal delay occurring between the pump and the
parametric photons varies in each portion of the whole path BBO-$M$-BBO.
Note that, because of the transmission through $\lambda -$qwp, SPDC
radiation and pump beam are respectively circularly and vertically polarized
when travel through the section between $M$ and $\lambda -$qwp. In order to
evaluate the total temporal walk-off, the following refractive indexes have
to be considered \cite{15}:
\begin{equation}
\begin{tabular}{|l|l|l|}
\hline
\ \ \ \ \textbf{Refractive index } & \textbf{BBO} & \textbf{Quartz} \\ \hline
$\ \ \ \ \ \ \ \ n_{o}(\lambda _{p})$ & $1.69355$ & $1.5585$ \\ \hline
$\ \ \ \ \ \ \ \ n_{e}(\lambda _{p})$ & $1.59574$ & $1.5681$ \\ \hline
$\ \ \ \ \ \ \ \ n_{o}(\lambda )$ & $1.6607$ & $1.5384$ \\ \hline
$\ \ \ \ \ \ \ \ n_{e}(\lambda )$ & $1.57013$ & $1.54725$ \\ \hline
\end{tabular}
\label{index}
\end{equation}
Note the different sign of birefringence occurring in BBO and quartz..After
a round trip, i.e. at the exit of BBO, $H-$polarized photons are delayed by $%
\tau =540f\sec $ with respect $V-$polarized photons. This value determines a
strong degradation of entanglement.

Concerning spatial walk-off, it is worth noting that, differently from the
type II crystal or the double crystal cases, because of the particular
geometry of our source, in the present condition the $\left|
H_{A}H_{B}\right\rangle $ and $\left| V_{A}V_{B}\right\rangle $ components
travel in two parallel directions with a negligible spatial displacement $%
\delta $ when go outside the BBO crystal for the second time.

The above considerations lead us to consider the relevant role of temporal
walk-off and discard spatial walk-off in the generation of polarization
entanglement. If temporal walk-off is not perfectly compensated the produced
state can be expressed by the density matrix:
\begin{equation}
\rho =\frac{1-v}{2}\left( \left| H_{A}H_{B}\right\rangle \left\langle
H_{A}H_{B}\right| +\left| V_{A}V_{B}\right\rangle \left\langle
V_{A}V_{B}\right| \right) +v\left| \Phi \right\rangle \left\langle \Phi
\right|  \label{mix}
\end{equation}
with $0\leq v\leq 1$ representing the weight of the entangled state
contribution. Temporal walk-off can be compensated by introducing an equal
amount of temporal delay of $\left| V_{A}V_{B}\right\rangle $ with respect
the $\left| H_{A}H_{B}\right\rangle $ component, restoring in this way the
temporal indistinguishability. In the next Section we'll examine in detail
some experimental results concerning this point.

\section{Purification by temporal walk-off compensation}

Temporal walk-of can be compensated by a suitable quartz plate with the
optical axis oriented in the $H$ direction in order to to recover temporal
indistinguishability of the $\left| H_{A}H_{B}\right\rangle $ and $\left|
V_{A}V_{B}\right\rangle $ components. Basing on the values of the ordinary
and extraordinary refractive indexes of (\ref{index}), $18mm$ of quartz are
necessary on this pourpose. The plate $Q$, shown in Figure 1, intercepting
the entire $e-ring$, was aligned at the output of the source with the optic
axis oriented along the $H$ direction.

We tested the polarization entanglement of the state $\left| \Phi
^{+}\right\rangle $ by using the two polarization analyzers ($\mathbf{\pi }$%
) shown in Figure 1 before the single photon detector modules ($A$ and $B$).
The experimental data given in Figure 3 show the$\ \pi $-correlation
obtained\ by rotating the first analyzer having kept fixed the axis of the
other at $45%
{{}^\circ}%
$. The interference pattern shows the high degree of polarization
entanglement of the source. The measured visibility of the coincidence rate,
$V$ $\simeq 90\%$, gives a strong indication of the entangled nature of the
state over an arc length of $2mm$, while the single count rates don't show
any periodic fringe behaviour, as expected. A polarization interference
visibility $\geq 85\%$ was measured over the entire $e-ring$.
\begin{figure}[h]
\centering
\label{fig3}
\includegraphics[scale=0.35]{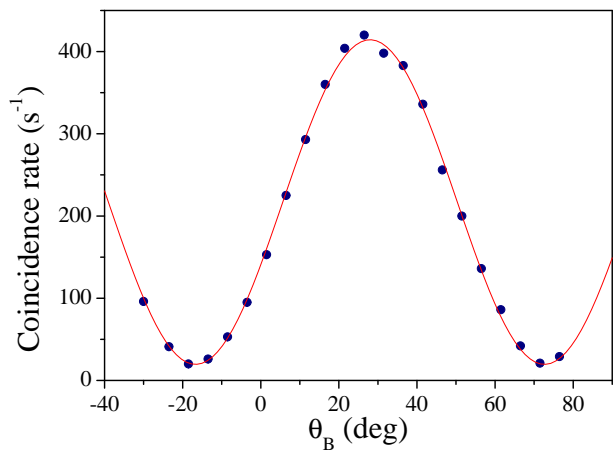}
\caption{Interference pattern of $\pi $-entanglement for the state $%
\left| \Phi ^{+}\right\rangle $, correspondong to a selected arc length of
the $e-ring=$ $2mm$.
}
\end{figure}

A further test of $\pi -$entanglement, after compensation of temporal
walk-off, is given by the tomographic reconstruction of the state $\left|
\Phi ^{-}\right\rangle $ shown in Figure 4.
\begin{figure}[h]
\centering
\label{fig4}
\includegraphics[scale=0.65]{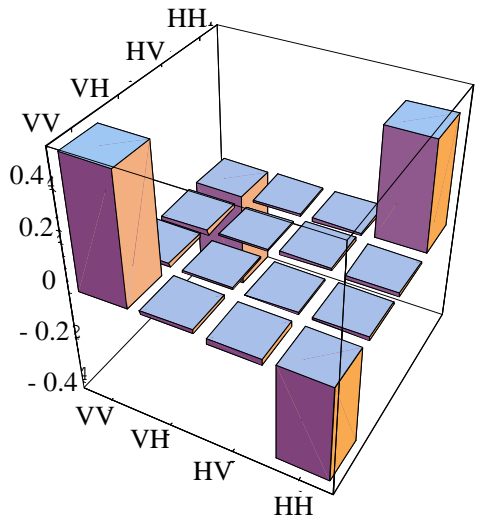}
\caption{Tomographic reconstruction (real parts) of the state
$\left| \Phi ^{-}\right\rangle$. The imaginary components are negligible.
Tangle $T=0.974\pm 0.091$, linear entropy $S_{L}=0.031\pm 0.059$, fidelity $%
F=0.949\pm 0.012$.}
\end{figure}

Finally, we performed a detailed investigation of the state $\rho $ of Eq. (%
\ref{mix}) by measuring the entanglement for different quartz plate
thicknesses. The experimental results referring to the entanglement
visibility $V$\ are summarized in Figure 5a and compared with the
theoretical predictions. These are obtained by following the theoretical
model of Ref. \cite{12} and calculating the convolution of the two photon
state amplitude with the gaussian shape of the $IF$\ transmission function.
As expected, the maximum visibility is obtained for a quartz thickness of $%
18mm$.

The entanglement witness operator is a method which allows to detect
entanglement with few local measurements \cite{16}. For the state $\rho $,
in the case of the state $\left| \Phi ^{+}\right\rangle $, it is given by
the expression
\begin{equation*}
W=\left( \left| \Psi ^{-}\right\rangle \left\langle \Psi ^{-}\right| \right)
^{PT}=\frac{1}{2}\left(
\begin{array}{cccc}
0 & 0 & 0 & -1 \\
0 & 1 & 0 & 0 \\
0 & 0 & 1 & 0 \\
-1 & 0 & 0 & 0
\end{array}
\right)
\end{equation*}
where $\left| \Psi ^{-}\right\rangle $ is the eigenvector corresponding to
the minimum eigenvalue of the partial transposition ($PT$) of $\rho $.

It is obtained
\begin{equation*}
Tr(W\rho )=-\frac{v}{2}
\end{equation*}
which is always negative for any value of $v\neq 0$. The experimental
results shown in in Figure 5b confirm the theoretical predictions, giving a
minimum value $W=-0.3808\pm 0.0043$ for a quartz thickness of $18mm$.

The last experiment, performed by varying the quartz thickness, consisted of
the measurement of the Bell-inequality parameter
\begin{equation}
S=\left| P\left( \theta _{A},\theta _{B}\right) -P(\theta _{A},\theta
_{B}^{\prime })+P\left( \theta _{A}^{\prime },\theta _{B}\right) +P\left(
\theta _{A}^{\prime },\theta _{B}^{\prime }\right) \right| \text{,}
\end{equation}
where
\begin{equation}
P(\theta _{A},\theta _{B})=\frac{[C\left( \theta _{A},\theta _{B}\right)
+C\left( \theta _{A}^{\bot },\theta _{B}^{\bot }\right) -C\left( \theta
_{A},\theta _{B}^{\bot }\right) -C\left( \theta _{A}^{\bot },\theta
_{B}\right) ]}{[C\left( \theta _{A},\theta _{B}\right) +C\left( \theta
_{A}^{\bot },\theta _{B}^{\bot }\right) +C\left( \theta _{A},\theta
_{B}^{\bot }\right) +C\left( \theta _{A}^{\bot },\theta _{B}\right) ]}\text{.%
}  \label{Bell-mom}
\end{equation}
and $C\left( \theta _{A},\theta _{B}\right) $ is the coincidence rate
measured by the two detectors and $\theta _{A}=0$, $\theta _{A}^{\bot }=\pi
/2$, $\theta _{B}=\pi /4$, $\theta _{B}^{\bot }=(3/4)\pi $. Figure 5c shows
the experimental results. The maximum value $S=2.465\pm .012$, corresponding
to a violation as large as $38$ standard deviations with respect to the
local realistic bound $S=2$, confirms once again the expected maximum for a
quartz thickness of $18mm$. It is worth noting that the violation of Bell's
inequality is a stronger condition than entanglement. Examples of non
separable states which admit a local realistic model can be given \cite{17}.
\begin{figure}[h]
\includegraphics[scale=0.25]{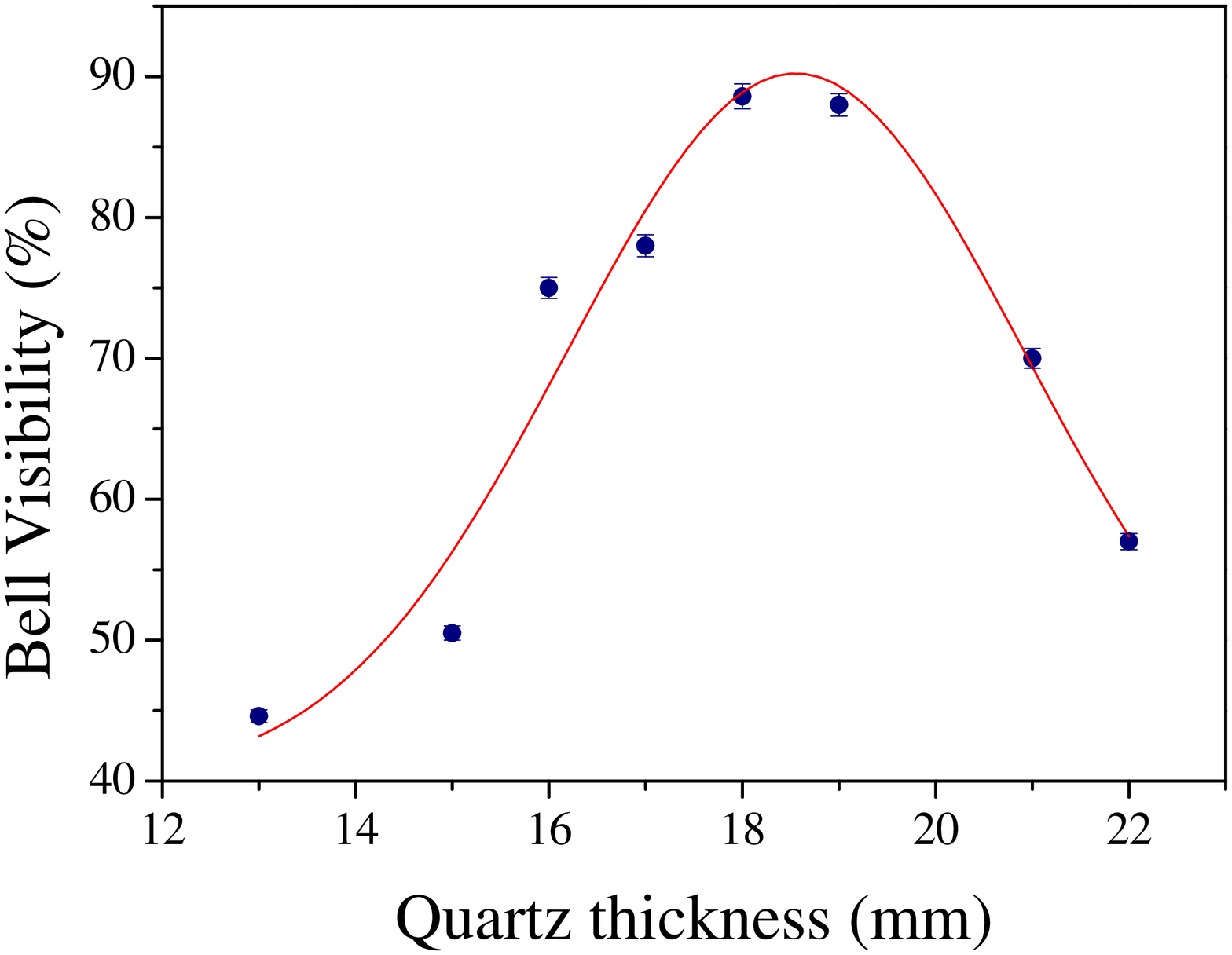}
\includegraphics[scale=0.25]{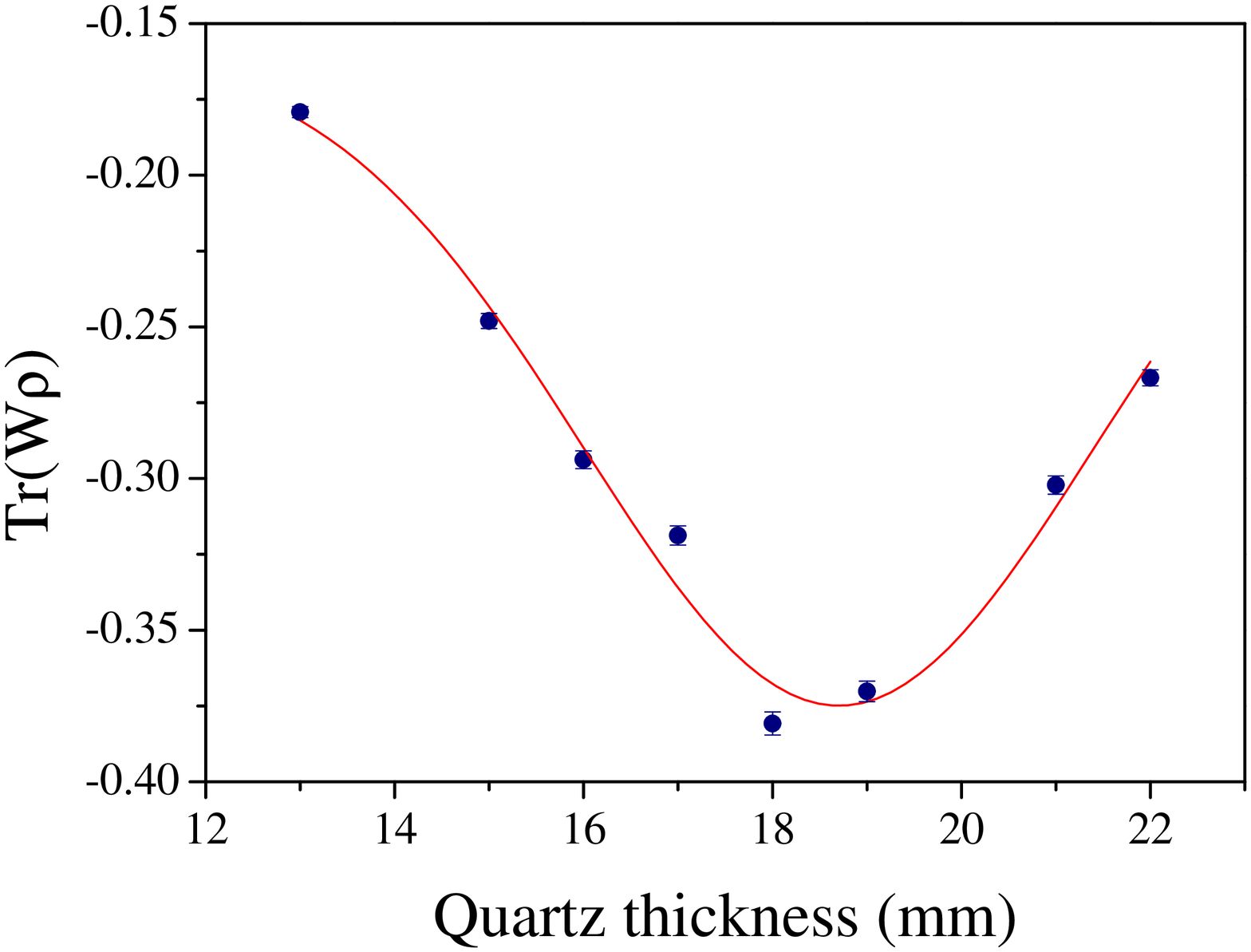}
\includegraphics[scale=0.25]{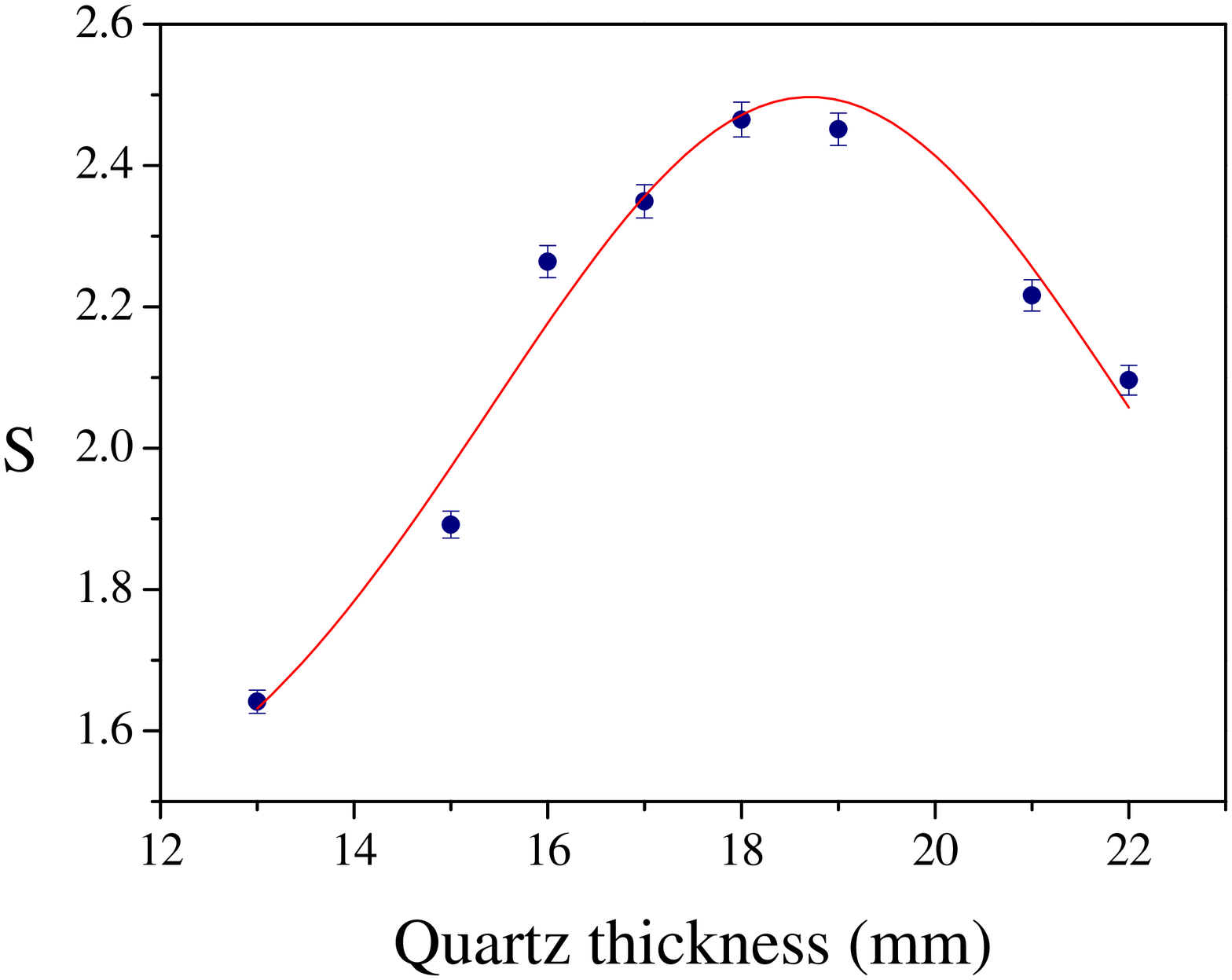}
\caption{a)  Entanglement visibilityof the state $\left| \Phi
^{+}\right\rangle $ measured for different values of quartz thickness.
b) Experimental results of entanglement witness for the
state $\left| \Phi ^{+}\right\rangle $, measured for different values of
quartz thickness.
c) Plot of the Bell parameter $|S|$ as a function of quartz
thickness.}
\end{figure}

\section{Conclusions and acknowledgments}

The structural characteristics of a single crystal parametric source of $\pi
-$entangled photon pairs, based on a single arm interferometer and operating
under femtosecond laser pulse excitation, has been described. We have
discussed the effects of temporal walk-off existing between the hortogonal
polarization components that, in these conditions, strongly affect the
purity of the entangled state. Polarization entanglement has been
investigated in a detailed way by measuring the state visibility, the
entanglement witness operator and the Bell nonlocal parameter $|S|$ as a
function of quartz thickness$.$ Highly pure entangled states has been
obtained by a suitable compensation of temporal walk-off.

This work was supported by the the FIRB 2001 and PRIN 05 of MIUR (Italy).


\begin{thebibliography}{99}
\bibitem{14}  The $\lambda -$qwp thickness has been evaluated by measuring
the temporal delay induced on the $f\sec $ laser pulse ($\lambda =795nm$) by
inserting it in one arm of a Michelson interferometer.
\end{thebibliography}
\end{document}